# PRODUCT CENTRIC HOLONS FOR SYNCHRONISATION AND INTEROPERABILITY IN MANUFACTURING ENVIRONMENTS

**Salah Baïna, Gérard Morel**

*CRAN UMR 7039, University Henry Poincaré Nancy I, F 54506 Vandoeuvre les Nancy, France,
{salah.baina, gerard.morel} @cran.uhp-nancy.fr*

Abstract: In the last few years, lot of work has been done in order to ensure enterprise applications interoperability; however, proposed solutions focus mainly on enterprise processes. Indeed, throughout product lifecycle coordination needs to be established between reality in the physical world (physical view) and the virtual world handled by manufacturing information systems (informational view). This paper presents a holonic approach that enables synchronisation of both physical and informational views. A model driven approach for interoperability is proposed to ensure interoperability of holon based models with other applications in the enterprise.

## 1. INTRODUCTION

Enterprise application integration and the opening of information systems towards integrated access have been the main motivation for the interest around systems interoperability. Integration aspect and information sharing in the enterprise lead to an organisation of the hierarchy of enterprises applications where interoperability is a key issue. In manufacturing enterprises, we can identify a hierarchy defines with three main levels:

- The higher Level represents management system level is responsible of the management of processes that handle all different informational aspects related to the enterprise (e.g.: ERP systems).
- The Execution level perform the processes that manage decision flows (e.g.: Workflow systems) and production flows (e.g.: MES);
- The Lower level is the process control level contains all processes that perform moves and physical transformations on the produced goods and services;

Our approach considers that the manufacturing enterprise is composed of two separated worlds rather than a simple hierarchy of levels :*(i)* On one hand, a world in which the product is mainly seen as a physical object, this world is called the manufacturing world it handles systems that are tightly related to the shop-floor level, *(ii)* On the other hand, a world where the product is seen as a service released in the market. This world is called the business world. (see Fig. 1):

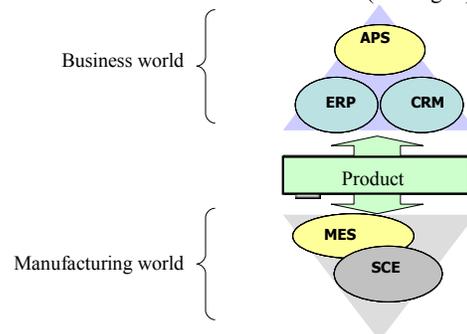

Fig. 1. Product centric approach

In order to achieve the main objective of the enterprise, "the product" to be specific, the business world and the manufacturing world need to exchange information and to synchronise their knowledge concerning the product (good and service). It is assumed that the product (good/service) can play the role of the gateway between both universes, since it represents a common entity between those worlds.

Hence, in this paper, we define a holon based approach in order to synchronise the business world and the physical manufacturing world using the holon concept. The paper continues by presenting a model driven based approach for ensuring interoperability between the holonic models and other models and representations of the enterprise.

Section 2 of the paper gives a brief overview of interoperability solutions in the enterprise. Section 3 presents the bases of our holonic modelling approach that uses the product as a centric entity in process models. Section 4 explains the model driven approach for ensuring holonic modelling interoperability with other modelling standards. In Section 5, an implementation of both modelling interoperability approaches is presented. Section 6 gives conclusions and perspectives for this work.

## 2. APPLICATIONS INTEROPERABILITY IN MANUFACTURING SYSTEMS

In this paper, the term "interoperability" refers to the ability to communicate, to cooperate and to exchange models or data between two or more applications despite differences in the implementation languages, the execution environments, or the models abstraction (Kalfoglou, and Schorlemmer, 2004). Interoperability problem has been studied in different domains. The following presents examples of existing methods and technologies enabling interoperability between enterprise applications.

### 2.1 ERP and MES Interoperability

The main purpose of interoperability between Enterprise Resource Planning (ERP) systems and Manufacturing Execution System (MES) is to improve the synchronisation between shop floor data and business information. ERP/MES interfacing speeds up the information flow between planning and process control systems, this allows tighter control of process timing and improvement of the quality of the production process. Defining this interface requires an information analysis to sort out reliable and accurate process data for the ERP system and to define proper recipe information for the MES. In the literature, several interoperability solutions between ERP and MES systems are proposed. Most of these solutions are based on message exchange technologies (e.g. BizTalk, XML messages, etc.). Even if proposed solutions resolve the problem of interconnection in a computer integrated manufacturing system, they are valid only for a specific ERP application and a specific production control system; they cannot be reused for other applications. Indeed in the context of interoperability paradigm, peer-to-peer solutions that create single links between many applications are known to be expensive in term of maintenance; the interoperability solution may become more expensive to maintain than the connected applications.

### 2.2 CRM and SCM Interoperability

Selk *et al.* (2005) present an integrative information system architecture for Customer Relationship Management system (CRM) and Supply Chain Management system (SCM). This architecture is based on a functional interconnection of both CRM and SCM components. On one hand, this approach improves collaboration between networked enterprises and accelerates information exchange between customers and suppliers, on the other hand it considers that all partners (customers and suppliers) have the same internal architecture which is not realistic; generally, even with high level of transparency each partner is free to choose the internal features of his information system.

### 2.3 Workflows Systems Interoperability

The workflow management coalition (WfMC) defined an interface for the communication and the interoperation of workflow engines. The interface defined in (WFMC, 1995) contains the specification of all API (Application Programming Interface) calls needed to cover all cases and possibilities when two workflow systems co-operate. WfMC interoperability standards and most of existing techniques that enable business process or workflow interoperability are based on a message exchange paradigm. These solutions resolve only the particular case of syntactic interoperability (messages vocabulary, format, data types).

In order to take into account interoperability requirements during the modelling phase in the context of manufacturing systems, we introduce, in the next section, the holonic process modelling approach based on the concept of holon. Afterwards, we will show how this modelling approach can interoperate with other models used in the enterprise.

## 3. A PRODUCT CENTRIC APPROACH FOR INTEROPERABILITY

Existing solutions for interoperability in enterprise environment focus mainly on enterprise processes interoperability and interconnection. Throughout product lifecycle, coordination needs to be established between the reality in the physical world where the product evolves as a physical object and the "electronic" world handled by manufacturing information systems where the virtual image of the product evolves as an informational object. Our work aims to provide a product centric approach for enabling interoperability between information systems in the manufacturing environment in order to establish the coherence between the physical products and their informational representations. To take into account this duality (physical things/ informational things), we propose an adaptation of the concept of holon to this specific problem.

*3.1 The Holon Concept in Manufacturing Process Modelling*

A holon is an identifiable part of a system that has a unique identity, yet is made up of sub-ordinate parts and in turn is part of a larger whole (Koestler, 1967). a Holonic Manufacturing System (HMS) is an autonomous and co-operative building block of a system for transforming, transporting, storing and/or validating information and physical objects. This concept has been widely used in manufacturing systems (McFarlane, and Bussmann, 2000; Van Brussel, *et al.*, 1998). In this paper, we adapt the *holon* concept definition to solve the problem of synchronisation between physical views and informational views of the same objects. We define the holon then as an aggregation of an information part and a physical part.

*3.2 The holon concept for product representation:*

In Holonic Process Modelling (Baïna, *et al.*, 2005; Morel, *et al.*, 2003), holons are used to represent products; the physical part of the holon represents the material part (also called physical view) of the product and the informational part of the holon represents the informational part (informational view) of the product. We consider that the information about a holon (a product) is distinguished into two categories:
- Information describing the current state of the holon, this state contains three kinds of data handling attributes of space, attributes of shape, and attributes of time (Panetto, and Pétin, 2005);
- Information that concerns the holon but does not correspond to any of the three types of attributes; space, shape or time. This information is defined by a set of properties (Morel, *et al.*, 2003).

Holons can be classified into two categories:

*(i) Elementary holons:* are the combination of a single informational part and a single physical part.

*(ii) Composite holons:* are the result of the processing and treatment of one or more other holons, this processing can be an assembling of existing holons, or a decomposition of a holon into several new holons. Each composite holon can be defined as the output of the execution of a manufacturing process on one or more less complex holons.

Fig. 2 represents the formalisation of holons in UML class diagram; here is a brief description of the classes defined in this diagram: The Class *Holon* defines basic attributes for both composite and elementary holons. An *Elementary Holon* is defined as a holon with no indication about his lifecycle. For example a product, produced by external manufacturing systems does not give information about the processes needed for his manufacturing. In general, an elementary state is observed and associated to each elementary holon. A *Composite Holon* is a holon that has been processed through at least a single process during his manufacturing. Only processes inside the domain of the enterprise are taken into account.

The *state* describes the current state a Holon (composite or elementary). Every manipulation of a holon through a process (Process Instance) implies a change in the state of the processed *holon*. A *Property* of a *holon* contains information that can not be handled only using its *state*. For the sake of traceability, it is possible to create a new state instance for each change in the existence of a *holon*, hence the whole lifecycle of the *holon* can be stored. An *Elementary state* describes specific data that concerns the state of an *elementary holon*. The *Composite state* class describes specific data that concerns the state of a *Composite Holon*.

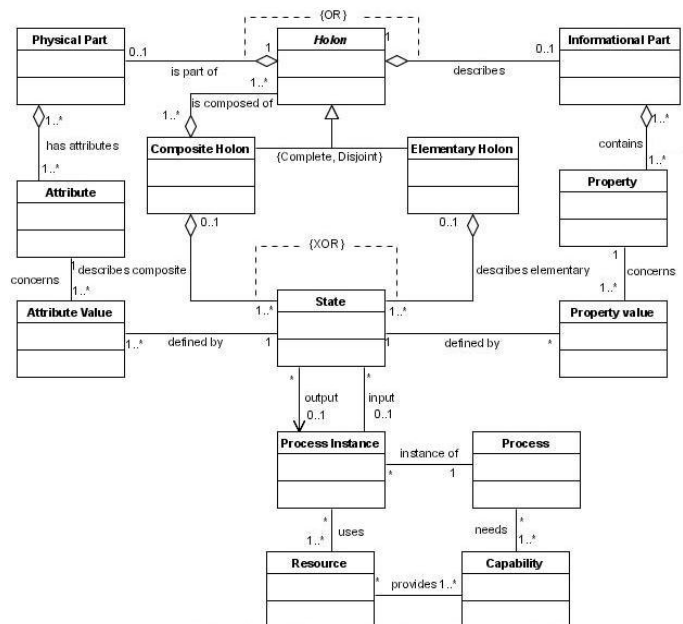

Fig. 2. Class diagram for the Holon model

A *Physical Part* is a reference to the physical part encapsulated in an *Elementary holon*. The *Process instance* refers to the execution of a process on a single *holon*, this class enables description of the execution of the process with high level of detail (e.g.: elapsed time, start and end of the treatment, used equipment, needed personal). A *Process instance* input is a *holon* state, this state can be composite or elementary but not both, hence a process instance cannot be associated to both a composite state and an elementary one. A *Process* describes an internal process that is performed inside the studied domain. The *Resource* class describes resources needed to perform a process instance. A *resource* can be either a material resource or a human resource.

In the next section, we will show how this approach can be used to express models based on the holon concept defined in section 3.2 using models related to existing data exchange standards and other unified languages. This transformation will enable model and data exchange between applications that are based on holon models and different enterprise applications from different levels in the enterprise.

## 4. MODEL DRIVEN APPROACH FOR HOLONIC MODELS INTEROPERABILITY

In this section, we introduce an approach for interoperability based in a model driven architecture (MDA) (Bezivin, 2001). The main objective of this section is to show how models based on the holon concept defined in section 3.1 could be expressed and transformed into models based on existing data exchange standards and other unified languages.

### 4.1 Interoperability in the MDA Context

Fig. 3 shows the four-level ontological approach levels for modelling that are used in the MDA. As it is explained in (Naumenko, and Wegmann, 2003), the lowest level $M^0$ presents different subjects for modelling, called universe of discourse. The level $M^1$ contains different models of each universe of discourse. The next level $M^2$ presents domain specific meta-models: one meta-model for each of the domains of interest relevant for the $M^1$ models. And finally, $M^3$ level presents a meta-meta-model designed to allow the definition of all the existing in the scope of the meta-models. In case of MDA, each application can be considered as a specific use of a model defined in the $M^1$ level which is based on meta-model defined in $M^2$. Application interoperability can then be resolved either by interconnecting applications together using a level $M^0$ exclusive reasoning, or by establishing a top-down approach for resolving applications interoperability based on the four levels of the MDA. Several research works have been done in order to resolve meta-models mapping problems. Lemesle (1998) explains how models transformation can be resolved by establishing transformation rules between meta-models. Transformation rules define a mapping that guides model transformations from the instances of the source meta-model to instances of the target meta-model.

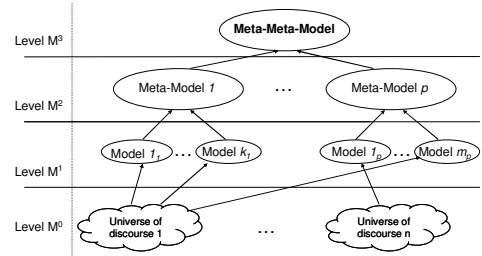

Fig. 3. The four-level ontological approach

In this architecture, to ensure interoperability between applications that handle those different meta-models, we should first define mappings that enable transforming one instance of a meta-model in an instance of another meta-model. Let us consider two applications A and B; A and B are interoperable, if and only if there is a mapping from the meta-model of A ($M_A$) to the meta-model of B ($M_B$) and a mapping form $M_B$ to $M_A$. Those mappings ensure that we can build a model compatible with A from a model used by B (and vice versa).

### 4.2 The MDA approach for Holon mappings

In Order to use the MDA approach for interoperability in the holonic context, we need to position the holonic modelling approach in terms of models meta-models and universe of discussion ($M^2$, $M^1$ and M0). In the holonic context, the universe of discourse $M^0$ concerns "Manufacturing enterprise", to describe this universe of discourse we use holonic models ($M^1$) that are instantiations of the meta-model defining holons and relationships with their context ($M^2$). The universe of discourse "Manufacturing enterprise" can also be described using other modelling approaches and languages dedicated to specific domains in the enterprise. To illustrate the use of mappings to interconnect the holonic approach to other models of the enterprise we chose two examples:
- The Unified Enterprise Modeling Language (UEML, 2003) used at the organisational level of the enterprise.
- The IEC 62264 standard (IEC 62264, 2002), which has been developed for interfacing the manufacturing control and execution systems with higher level systems.

Both UEML and IEC 62264 are elements in the $M^2$ level of the MDA hierarchy. On one hand, UEML represents the link between the holonic representation of the product and the business universe of the enterprise, on the other hand the IEC 62264 interface linking this holonic representation to the manufacturing environment.

Mapping Holon with UEML. The Unified Enterprise Modelling Language (UEML) is the result of the UEML project (UEML, 2003). The UEML is an Interlingua between Enterprise Modelling tools. The meta-model of UEML1.0 defines the set of most relevant concepts and notions for Enterprise modelling. In the following, a simple mapping is presented. This mapping consists in representing

*Informational part* (resp. *Physical part*) using the concept *Information Object* (resp. *Material resource*), to represent holons we need a specific sub-type of the class *Object*, this specific concept must be the link between the *Material resource* (*Physical part*) and the *Information Object* (*Informational Part*).

In the UEML meta-model, an activity represents a generic description of a part of enterprise behaviour that produces outputs from a set of inputs. It represents a set of similar activities executions. The concept of "Activity" defined by the UEML constructs can be used to express the notion of "Process" which is defined in the holon model. Table 1 summarises the mapping between concepts defined in the holon model and UEML constructs.

*Table 1: Correspondence between Holon concepts and UEML concepts.*

| Holonic Process Model Meta-model | UEML Meta-Model |
|---|---|
| Holon | Object |
| Informational Part | Information Object |
| Physical Object | Material resource |
| Process | Activity |

Mapping with the IEC 62264 standard. The IEC 62264 is the standard specifying the exchange of data and models interfacing the shop floor level into the enterprise (IEC 62264, 2002). It is composed of six different parts designed for defining the models and interfaces between enterprise activities and control activities. Each model concerns a particular view of the integration problem. Those models show increasing detail level in the manufacturing system. They can be classified in to two categories; operational models or resource models. To map the holon concept with the IEC 62264 models, two views are possible:

*Genealogical view*: it concerns the genealogy of the holon, which describes relationships between holons (composition and transformation). Information handled in this view corresponds to the "Product Definition Information", as defined in the standard;

*Process view*: it allows the representation of the lifecycle of the holon and the different processes involved during each step in the manufacturing cycle. This view fits into the "*Product production rules*" concept of the IEC 62264 standard, especially the "*Product segment*" concept, which can manage information about assembly steps and assembly actions for discrete manufacturing.

The IEC 62264 material model is the most adequate to express and handle the information about genealogy of products (Baïna, *et al.*, 2005); it is used to represent the genealogical view of a holon. In the IEC 62264, the notion of product segment defines the values needed to quantify a segment for a specific product, such as the number with specific qualifications. In the holon model in Fig. 3, the class "*Process instance*" groups all information that concerns the manufacturing of a specific holon. Since we consider that a holon represents a product, we can then assume that the "*Process instance*" of a specific holon describes the product segment of that holon (product). This means that all concepts associated to the *Process instance* class can match the concepts associated to Product segment. Table 2 shows the mapping between the genealogical aspect and the process view of the holon model into the material model of IEC 62264 standard. To implement the mapping between holons and IEC 62264, we use the Business to Manufacturing Mark-up Language (B2MML, 2003), which is an XML implementation of the IEC 62264 part 1.

## 5. MODELLING TOOLING

### 5.1 Holonic models instrumentation

To experiment the holonic approach (modelling and mappings) defined above in a real case, we integrate it into a commercial CASE tool named MEGA (http://www.mega.org). MEGA suite is an enterprise process modelling tool that contains a business process analysis tool, modelling tools and design environments. MEGA has its own meta-model that described all concepts and objects ready to use, and all relationships that exist between those concepts. This meta-model can be customized and specialised for specific users' needs. MEGA Suite can be used to define, describe and exploit several kinds of diagrams (e.g: Business process Diagrams, UML Diagrams, Workflows). In our contribution, we focus only on business process diagrams; indeed they seem to be the most adequate choice for holon integration. Business Process diagrams in MEGA are based on a meta-model very similar to the Business Process Modelling Notation of the OMG (http://www.bpmn.org).

Fig. 4 shows part of MEGA meta-model describing the most important concepts contained in business process diagrams. MEGA offers tools that enable customizing the meta-model; we used these tools to embed our own holon model into the meta-model of Mega in order to test the usability of our proposal.

*Table 2: Correspondence between Holons related concepts and IEC 62264 concepts*

| | Holonic Process Model | IEC 62264 |
|---|---|---|
| **Material Model mapping** | Holon | Material sublot |
| | Holon Flow | Material lot |
| | Informational Part | Material definition |
| | Properties and Attributes | Material lot property definition |
| **Product Definition Model mapping** | Process Instance | Product segment |
| | Equipment | Equipment specification |

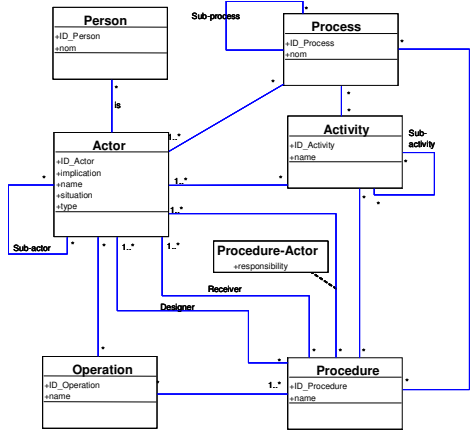

Fig. 4. Meta-model of MEGA business process diagrams

In addition to the meta-classes defined in the meta-model shown in Fig.5, other meta-classes exist; one of the most important is the meta-class MESSAGE that describes the messages exchanged between objects and entities involved in a system. First, to implement our approach, we replace the concept of message by a more general concept: flow. Flows can be from three types; holon flows that contain holons, informational flows that contain informational parts and physical flows that contain physical parts. We integrate the concept of flow in the Business Process Diagram meta-model as a meta-class. After the flow meta-class we define all other classes related to this concept (Baïna, *et al.*, 2005). We integrated also property and attribute notions that describe information related to holons.

In the next part of this section, we briefly explain how information contained in diagrams that have been modelled using the new features and concepts integrated in MEGA is extracted and mapped into other standards in order to be reused by other applications.

*5.2 Mappings implementation*

To implement the mappings from the holonic models designed in MEGA and the other formats, we chose the XML format to represent data extracted from our holonic models. A customized XML generator enables generating different XML files with specific structures; each file corresponds to a specific standard or language. B2MML and UEML XML schemas are used to generate valid documents based on the mappings defined in section 4.2, those documents can be imported in applications compatible with UEML or B2MML formats. The translation from the XML generated directly from the holonic diagrams, to the different XML schemas is based on XSLT transformation rules that implement the mappings defined in section 4.2 to transform an XML input file describing a holonic model to generate an XML output file in UEML or B2MML formats.

## 6. CONCLUSION

In this paper, we defined an approach for specifying holonic Process Models. This concept enables maintaining synchronisation between the physical objects and their informational views. We propose an enterprise interoperability approach based on a model driven architecture applied to holonic models. This approach uses mappings that transform holonic models into UEML and IEC 62264 structures. Finally, we presented an implementation of the whole approach in a commercial modelling tool. Perspectives for this work concern the formalisation of the mappings in terms of enterprise ontologies and formalised ontology mappings.

## REFERENCES

B2MML (2003) The World Batch Forum. Business To Manufacturing Markup Language, version 2.0, 2003.

Baïna, S., H. Panetto and G. Morel (2005). A holonic approach for application interopearbility in manufacturing systems environment. *In Proc of the 16th IFAC World Congress, Prague, Jul 2005*

Bezivin, J. (2001). From Object Composition to Model Transformation with the MDA. *Proceedings of TOOLS* USA, Santa Barbara, August 2001.

IEC 62264 (2002). IEC/ISO FDIS 62264-1:2002. *Enterprise-control system integration, Part 1. Models and terminology*, IEC and ISO, Geneva.

Kalfoglou, Y. and M. Schorlemmer (2004) Formal Support for Representing and Automating Semantic Interoperability. *In ESWS'04, The 1st European Semantic Web Symposium*, Heraklion, Greece.

Koestler A. (1967) The Ghost in the Machine Arkana, London.

Lemesle, R. (1998). Transformation Rules Based on Meta-Modelling. EDOC'98, California, USA, Nov 98.

Mc Farlane, D and Bussmann, S. (2000) Developments in holonic production planning and control, *International Journal of Production Plannig and Control*, **Vol. 11, N° 6**, pp. 522-536

Morel G., H. Panetto H., M.B. Zaremba and F. Mayer (2003). Manufacturing Enterprise Control and Management System Engineering: paradigms and open issues. IFAC Annual Reviews in Control. **27/2**.

Naumenko, A., A. Wegmann (2003). Two Approaches in System Modelling and Their Illustrations with MDA and RM-ODP. In *ICEIS 2003, the 5th International Conference on Enterprise Information Systems*.

Panetto H. and J.F. Pétin, (2005). Metamodelling of production systems process models using UML stereotypes, *International Journal of Internet and Enterprise Management*, **3/2**, 155-169

Selk, B., S. Kloeckner and A. Albani (2005). Enabling interoperability of networked enterprises through an integrative information system architecture for CRM and SCM, *In BPM 2003, the 3rd Business Process Modelling Conference, Nancy, Sept 2005*.

UEML. (2003). Unified Enterprise Modelling Language (UEML) Thematic Network. IST-2001-34229

Van Brussel, H, J, Wyns, P, P, Valckenaers, L, Bongaerts and P, Peeters (1998). Reference Architecture for holonic manufacturing systems: Prosa. *Computers in Industry*. **37 (3)**, pp. 255-274

WFMC (1995). Workflow Management Coalition: Workflow standard – Interoperability abstract specification. Document Number WFMC-TC-1012.